# Effect of the build orientation on mechanical and electrical properties of pure Cu fabricated by E-PBF

Alizée THOMAS[1,2], Guillaume FRIBOURG[2], Jean-Jacques BLANDIN[1], Pierre LHUISSIER[1], Rémy DENDIEVEL[1], Guilhem MARTIN[1,#]

1. Univ. Grenoble Alpes, CNRS, Grenoble INP, SIMAP, F-38000 Grenoble

2. Schneider Electric, Technopole, 28 rue Henri Tarze, Grenoble, France

#corresponding author: guilhem.martin@simap.grenoble-inp.fr

## Abstract

Electrical conductors are usually made of pure copper because this material shows outstanding electrical conductivity (IACS > 100%). Additive manufacturing can be used to further improve the performances of such electrical conductors because it enables to consider sophisticated geometries impossible to be fabricated by conventional processing routes such as casting, machining or hot forming. In the present work, Electron Beam Powder Bed Fusion is optimized to achieve dense pure copper specimens (relative density >99.9%). Only few residual spherical pores inherited from the as-received powder particles are detected in the samples as nicely revealed by X-ray microtomography scans. Three different build orientations, namely 0°, 45° and 90° with respect to the build direction are produced. The microstructures are characterized using optical microscopy and EBSD measurements. The electrical conductivity is measured using the Eddy current method as well as the four-probes method while the mechanical properties are assessed using Vickers microhardness and tensile mechanical testing. The electrical and mechanical performances are discussed in the light of the microstructural characterizations, and further compared with a cold-worked Cu-ETP sheet subjected to an annealing treatment. It demonstrates that E-PBF is able to reliably produce components made of pure Copper whose properties equal those of pure copper fabricated by more conventional processing routes, regardless of the build orientation.

**Keywords:** Electron Powder Bed Fusion; Copper; microstructure; mechanical properties

# 1. Introduction

Pure copper exhibits remarkable physical properties with a thermal conductivity of 401 W/m.K and an electrical conductivity of 100%IACS (58 MS/m) in the annealed conditions. On top of that, copper shows a good corrosion resistance and antibacterial properties. All those properties make copper a good candidate for a variety of applications in the energy sector: heat exchangers, electrical conductors [1]. There is recently a growing interest in fabricating copper components by Additive Manufacturing (AM) because such processing routes offer various advantages, as compared to conventional manufacturing processes. First, AM is considered as an efficient way to produce prototypes which speeds up the development of innovative components. Second, it allows the production of customized parts with sophisticated geometries [2] [3]. Third, the optimization of the design of parts made of pure copper can lead to a reduction of raw material consumption. Finally, optimized designs are expected to improve their performances, especially their energy efficiency. Thus, it turns out to be crucial to master the manufacturing of pure copper conductive parts using AM.

Among AM-processes, L-PBF (Laser Powder Bed Fusion) and E-PBF (Electron Powder Bed Fusion) are certainly the ones offering the best guarantees for fabricating complex geometries such as heat exchangers with internal channels or electrical conductors integrating architectured materials. Producing dense copper parts using L-PBF with conventional IR-lasers (wavelengths in the range 1000-1100nm) is very challenging because of its extremely high reflectivity that usually leads to porous samples (density <98%) and molten pool instabilities [4] [5] [6]. Less than 5% of the incident laser energy is absorbed by the material while the major part is reflected [7]. Very recently, two avenues have been explored to overcome this challenge of processing dense pure copper components using L-PBF. The first option consists in applying very high laser power, typically more than 500 W to generate more stable molten pools and transfer more heat to the material [8] [9] [10]. This leads to improvements on the density achieved after L-PBF but this could decrease the lifetime of some machine components (lenses), if some precautions are not applied. The second option consists in using green/blue lasers that show an increased absorptivity (in the range 20 - 60%, less than 5 % for IR-lasers) [11] [12].

On the contrary, it was demonstrated that E-PBF is a suitable process to achieve relative densities higher than 99.5%. This was attributed to the high absorptivity of the beam energy when using electrons. In addition, E-PBF works under vacuum, a suitable environment to prevent parts contamination that could eventually further lead to a reduction of the electrical conductivity. Several studies have reported successful attempts to produce dense pure copper components by E-PBF [2], [13], [14], [15] , [16], [17], [18], [19]. However, the relationships between the microstructure and the mechanical and functional properties have only been partially addressed. Very few data have been published regarding the effect of the build orientation on the tensile response of pure copper parts fabricated by E-PBF. One of the published articles, see [18] shows surprising results with drastically contrasted tensile responses depending on the build orientation: tensile specimens built with their loading axis perpendicular to the build direction turn out to be brittle. Overall, there was not a lot of efforts devoted to discuss the measured mechanical properties. Moreover, the data available in the literature regarding the electrical conductivity were measured using the Eddy current method which suffers from a major issue: it only reflects the conductivity of a small volume of material underneath the probe [16] [17]. Thus, there is a need to assess the electrical conductivity using a method that actually provides a bulk measurement, much more representative to reflect the behaviour of an electrical component. Such pieces of information are crucial to demonstrate the ability of E-PBF to reliably produce dense components with optimized properties.

The objective of this work is to establish relationships between microstructures and both mechanical (tensile behaviour) and functional (electrical conductivity) properties of Cu-parts fabricated by E-PBF. Dense specimens built in three different orientations with respect to the build direction are first produced with an optimized set of processing parameters. The microstructure (grain size and morphology) and microtexture of the samples built with different orientations are characterized using metallography and SEM-EBSD. The mechanical and electrical properties of the parts built with various orientations by E-PBF are discussed in the light of their microstructure and further compared with the properties of a Cu-ETP produced with more conventional processing routes (wrought product).

## 2. Experimental procedures

### 2.1. Materials and E-PBF fabrication

The copper powder was produced by gas atomization (argon) from Cu-OF rods with 25 ppm of oxygen using the powder atomization facility of an academic partner, namely the LERMPS Laboratory in France, a department of the UTMB (Université Technologique de Belfort-Montbéliard) [20]. The composition of the copper powder meets the requirement for Cu-ETP grade with a purity of 99.93 %wt in Cu. The detailed chemical composition of the as-received powder batch is given in **Table 1** and was done by Electrowerk Weissweiler GmbH [21].

| Cu | H# | O* | P | B | Mg | Al |
|---|---|---|---|---|---|---|
| 99.93 | 0.0001 | 0.011 | < 0.005 | 0.002 | < 0.003 | 0.001 |
| **Si** | **S** | **Ca** | **Ti** | **V** | **Cr** | **Mn** |
| < 0.005 | 0.002 | 0.003 | 0.006 | < 0.001 | < 0.003 | < 0.001 |
| **Fe** | **Co** | **Ni** | **Zn** | **As** | **Se** | **Mo** |
| < 0.005 | < 0.003 | < 0.003 | < 0.003 | < 0.003 | < 0.003 | 0.001 |
| **Ag** | **In** | **Sn** | **Sb** | **Te** | **Pb** | **Bi** |
| < 0.002 | < 0.010 | < 0.003 | < 0.001 | < 0.003 | < 0.005 | < 0.003 |

**Table 1.** Chemical composition of the as-received powder batch in %wt. #: measured by Thermal Conductimetric method after fusion in a current of inert gas. *: measured by Infrared method after fusion under inert gas. All the other elements were measured by Inductively Coupled Plasma optical spectrometry.

The as-received powder morphology was characterized using a ZEISS GEMINI SEM500 field emission gun scanning electron microscope (FEG-SEM). As shown in **Figure 1**, the as-received powder mostly exhibits a spherical morphology but with numerous agglomerates (see example pointed out with a yellow arrow in **Figure 1**) and satellites. The particle size distribution was measured by laser diffraction (Malvern Mastersize 2000) giving a $Dv_{50}$ equal to 70 µm. The relative apparent density and flowability were also measured based on the ASTM B212 (standard test method for the relative apparent density of free-flowing metal powders using the Hall flowmeter funnel) and ASTM B213 (standard test methods for the flow rate of metal powders using the Hall flowmeter funnel) respectively. The relative apparent density was measured to be 45.5% and a flow time of 17s was found.

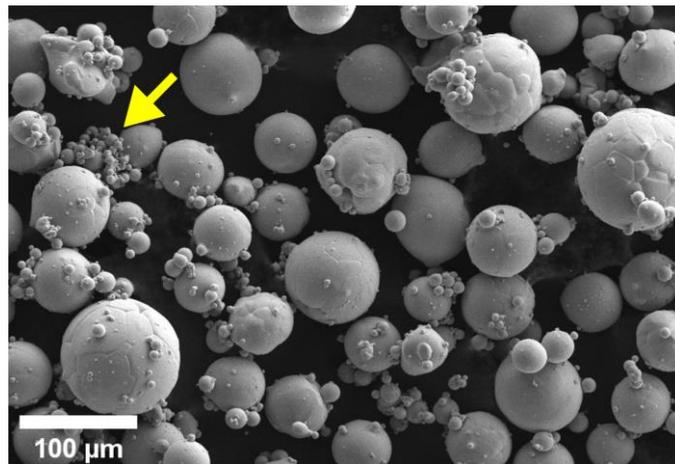

**Figure 1.** (a) Powder morphology as-revealed by SEM under secondary electron contrast (SE).

Powder was loaded into an ARCAM A1 E-PBF machine from operating under vacuum ($2.10^{-3}$ mbar) and at an accelerating voltage of 60 kV. A home-made miniaturized building chamber that fits into the ARCAM-A1 standard building chamber was used in this study in order to reduce the amount of powder used for a build. $70{\times}70{\times}10$ mm$^3$ start plates made of Cu-ETP were used. It is important here to highlight that due to its physical properties (electrical conductivity of 58MS/m and density of 8.84g/cm$^3$), Cu is

| Preheating area (mm²) | Beam Current (mA) | Beam Speed (mm/s) | Focus Offset (mA) | Number of repetitions | Line Order | Line Offset (mm) |
|---|---|---|---|---|---|---|
| 60 × 60 | 6 | 11 000 | 62 | 9 | 10 | 0.1 |

**Table 2.** Preheating parameters used to produce the pure Cu parts in our in-house miniaturized build chamber.

A focused beam was used to melt the powder bed with a snake-like strategy, adjacent molten tracks being spaced by 0.1 mm. The layer thickness was set to 50 μm and the scanning direction was rotated by 90° at every layer. The automatic mode of the software that controls the beam parameters was disabled. This is important because when the automatic mode is turned on, different functions are activated to control the electron beam to improve the build quality for sophisticated geometries by adjusting the beam parameters (power, scan speed) to locally vary melting conditions as a function of the scan lengths or the presence of negative surfaces. Bar samples with the largest dimension being equal to 55 mm were built using a set of parameters [v = 3000 mm.s⁻¹; P = 1200 W] that was preliminary optimized to maximize the final density of the built parts [22]. The preliminary optimization was based on a design of experiments exploring beam power ranging from 900 to 2500 W and scan speed ranging from 1000 to 5000 mm/s. The set of parameters employed to produce the samples in this work is consistent with the processing window identified in a recent publication, see [19]. Three different build orientations, namely 0, 45 and 90° with respect to the build direction were fabricated with a constant scan length in both scanning directions (x and y) of 10 mm, see schematic side and top views in **Figure 2a** and **Figure 2b** respectively. The 90°-specimens were positioned diagonally on the build substrate to maintain a scan length close to 10 mm. Keeping the scan length constant and equal to 10 mm was deliberately chosen to achieve similar melting conditions regardless the build orientation. Three builds were produced: one per sample orientation (see example for the 45°-build in **Figure 2c**) and only fresh powder was used for the fabrication of the samples in order to avoid any issue related to powder recycling. Samples were built on 1 mm-diameter and a 3 mm-height cylindrical supports.

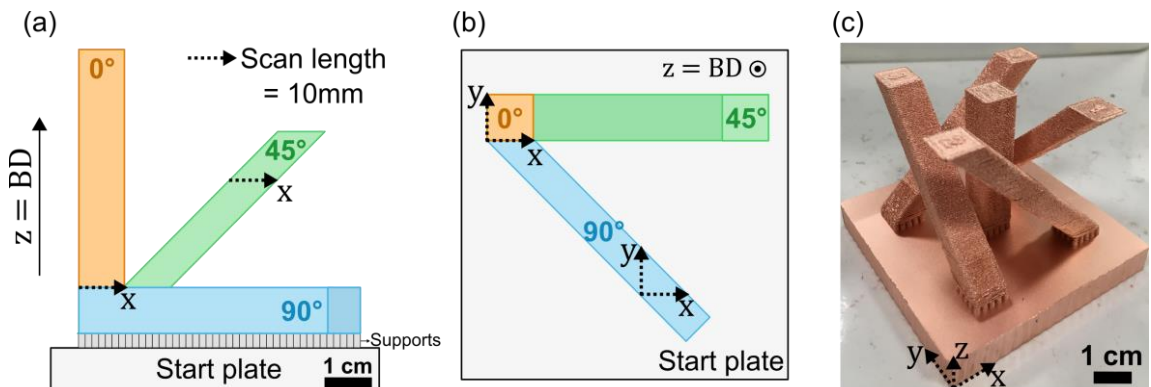

**Figure 2.** Schematic illustration of the bar sample orientations on the start plate: (a) side and (b) top view. (c) 45° bar samples built on a 70x70x10 mm³ copper start plate by E-PBF. (x,y,z) is the coordinate system of the start plate, x and y being the scanning directions, and z the build direction (BD).

## 2.2. Microstructural characterization

Samples were prepared for optical microscopy: grinding from 800 down to 2400 using SiC abrasive papers and polishing with 3 and 1 μm diamond suspensions. Due to the high ductility of copper, porosity in the sample can be filled in during grinding and polishing. Thus, an additional polishing step was carried out using a solution consisting of 100 mL $H_2O$, 100 ml ethanol and 19 g of iron (III) nitride as lubricant. This chemical treatment helps to clear out the pores as well as to reveal the grain boundaries. However, as such a chemical polishing reveals also the grain boundaries, estimating the porosity required a final gentle polishing using the 1 μm diamond suspension. The latter operation allowed the etching effects to be removed without filling the pores. Optical micrographs were then taken with an Olympus DSX 510 opto-numerical microscope. Porosity was estimated using image analysis by thresholding optical micrographs taken in the xz-plane. Area of $10 \times 10$ mm$^2$ were analyzed for each sample to ensure statistical consistency. To evaluate the uncertainties of such a measurement, two additional thresholds were manually chosen, a minimum and a maximum around the value sets by the automatic threshold.

As the estimation of porosity using image analysis based on optical micrographs might suffer from some limitations (2D method: pores underneath the surface are not included in the analysis), the porosity in both the as-received powder and samples fabricated by E-PBF was further characterized using X-ray microtomography. A Easytom-XL lab tomograph from RX-solutions® was used with an X-ray source Hamamatsu L10711 transmission tube with a LaB6 filament and a tungsten target (1 μm thick). The detector is a flat panel Varex 2520, with $1920 \times 1536$ square pixels of 127 μm. The acquisition and reconstruction parameters are summarized in **Table 3**. With a voxel size of 0.7 μm, objects with a diameter smaller than 1.5 μm cannot be detected.

| | |
|---|---|
| **Voxel size (μm)** | 0.7 |
| **Tube voltage (kV)** | 100 |
| **True target current (nA)** | 10 |
| **Filter** | 0.25 mm Al |
| **Frame rate** | 3 |
| **Number of average frame** | 7 |
| **Number of projections** | 1120 |
| **Scan range (°)** | 360 |
| **Acquisition time (min)** | 45 |
| **Scan height (mm)** | 1.1 |

**Table 3.** X-ray tomography acquisition and reconstruction parameters.

To characterize the raw powder, a 250 μm diameter glass capillary was filled with powder and glued to a ceramic sample holder. Cylindrical specimens with a 0.5 mm-diameter and their revolution axis parallel to the build direction were also extracted from the as-built samples by electron discharge machining (EDM) and subjected to X-ray microtomography scanning. The X-Act software was used to reconstruct the volumes via filtering the back-projection and a beam hardening correction. The 3D images were further analyzed with Fiji®. Volumes were filtered with a 3D median filter with a radius of 1 voxel. An automatic threshold was applied to discriminate the pores from the bulk material. To analyze the scan acquired on powder particles, the particles were first isolated from each other: the internal pores were first filled applying first the "3D Fill pores" algorithm followed by the application of a "3D Watershed Split" procedure available in Fiji®. The "Analysis 3D" plugin was finally applied to determine the size distribution of pores and particles [23].

For Scanning Electron Microscopy (SEM) observations, samples were mechanically polished down to 1 μm with a diamond suspension, and electropolished in a 65 % aqueous phosphoric acid solution. The electropolishing conditions were set to 5 V and 2 mA for 120 seconds with a stainless-steel cathode. SEM images were acquired with a ZEISS GEMINI SEM500 with an operating voltage of 15 kV. For EBSD analysis, a step size varying from 2 to 5 μm was employed. Electron Back-Scattered

Diffraction (EBSD) maps were post-treated using the OIM Analysis software. The grain widths were measured using the intercept method with lines drawn perpendicularly to the build direction.

### 2.3. Mechanical characterization

Microhardness measurements were performed on polished as-built samples using a Wilson Tukon 1212 Vickers Tester with a load of 500 g (HV0.5). For each sample, 18 measurements (3 lines of 6 measurements) along the build direction were performed. The error was estimated by computing the standard deviation over the 18 measurements.

Dog bone tensile specimens were extracted by EDM from the as-built bar samples as schematically illustrated in **Figure 3a**. Their geometry and dimensions are given in **Figure 3b**. A coordinate system attached to the tensile specimens is defined (TrD, ThD, LD), TrD being the transverse direction, ThD the thickness direction, and LD the loading (tensile) direction. Tensile specimens with the same geometry and with the tensile direction aligned with the rolling direction (RD) were also extracted from a work-hardened Cu-ETP sheet annealed at 900°C during 15 min for comparison with the samples produced by E-PBF. Such an annealing leads to a fully recrystallized microstructure. Tensile tests were conducted at room temperature using a DY35 testing machine with a 5 kN load cell and a displacement rate of 1 mm/min. Digital Image Correlation (DIC) was employed to measure the strain throughout the test. The DIC technique consists in recording the tensile test with two cameras. 6 Mpixels cameras with 100 mm lenses and GOM Correlate Software were used. The facet size was set to 20 pixels and the distance between two facets to 16 pixels. The samples were painted 1 hour before conducting the test. First, the samples were covered in white with a spray painting. Then, a fine black random speckle pattern was created with an IWATA airbrush. This pattern allows the image correlation algorithm to track the position of several surface points discriminated by their pixel intensity. The algorithm is able to measure the local displacement of each point by comparing two successive images. It allows the local strain to be calculated to produce strain maps. Besides local extensometers can be defined in various directions to measure the length changes between two points. At least three specimens for each orientation were deformed under tension up to fracture.

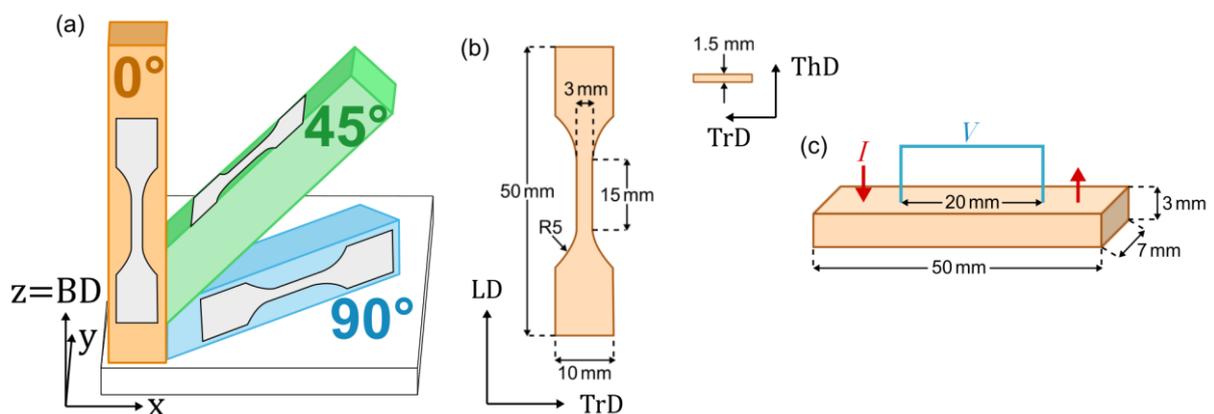

**Figure 3.** (a) 3D schematic illustrating how the samples with different build orientations were fabricated and indicating how tensile specimens were extracted from the bars. Dimensions of (b) the dog bone tensile specimens, and (c) specimens for electrical conductivity (4 probes method). (x,y,z) is the coordinate system of the start plate, x and y being the scanning directions, and z the build direction (BD). (TrD, ThD, LD) is the coordinate system attached to the tensile specimens.

### 2.4. Electrical conductivity measurements

The electrical conductivity of the samples produced by E-PBF was measured using two different methods: (i) the Eddy current (surface measurements), and (ii) the 4 probes method (bulk measurements). Parallelepipedic specimens with the geometry and dimensions indicated in **Figure 3c** were machined by EDM from the as-built bar samples.

The first method consists in applying an alternative magnetic field with a frequency $f$ using a probe in contact with the sample to create the so called Eddy current perpendicularly to the axis of the probe. This method is straightforward and can be applied on relatively small samples. However, the penetration depth is low, especially for high conductive samples, about 100 µm for dense copper having a 100%IACS electrical conductivity. Thus, this method provides only a measurement of the electrical conductivity of a small volume underneath the probe. A Fisherscope MMS PC2 Sigmascope equipped with a 3 mm diameter probe was employed, the frequency was set to 480 kHz. The probe was applied five times on a polished surface of the sample and the electrical conductivity was taken as the average of those five measurements. The error was then estimated by computing the standard deviation.

The second method relies on the four probes method. A continuous current $I$ is applied to the sample of section $S$, between two probes and the voltage $V$ is measured between two other probes separated by a distance $L$. The electrical conductivity $\sigma_{el}$ is then deduced from equation (1).

$$\sigma_{el} = (I.L)/(VS)$$ (1)

Six measurements were done using different currents: 1, 2, 4, 6, 8 and 10mA. Note that the electrical conductivity is very sensitive to the section measurement. The section $S$ was therefore measured with an electronic sensor Micromaster Capaµm System with an accuracy of ± 0.5 µm which leads to about 0.2% measurement uncertainty on the electrical conductivity.

## 3. Experimental results

### 3.1. Density and defects

The optical micrograph of a copper sample built vertically (0°-sample) by E-PBF is shown in **Figure 4a**: no lack-of-fusion neither chimney pores are observed. Similar observations were carried out for the two other build orientations giving a relative density estimated by image analysis of 99.95 ± 0.02% regardless the build orientation. Such a relative high density is consistent with the results reported in previous studies, see e.g. [17], [18], [19] indicating relative densities >99.5% with the best samples achieving 99.95% [16]. In such an optical micrograph, only one type of defects is observed: spherical pores as pointed out by the blue arrow in **Figure 4a**. Interestingly, such spherical pores were detected in cross sections of the as-received powder as exemplified in **Figure 4b**. A cylindrical sample with a diameter of 0.5 mm and extracted from a 0°-sample was further characterized by X-ray microtomography using a voxel size of 0.7 µm to have a better picture of the size and spatial distribution of those spherical pores in the built samples. In the 3D reconstructed image shown in **Figure 5a**, very few spherical pores are detected, the relative density is measured to be about 99.96 ± 0.02 % (0.04% porosity) confirming the preliminary optimization of the processing parameters conducted prior to this study and the estimation made by image analysis. Those spherical pores are randomly distributed in the sample and exhibit a sphericity of about 0.9.

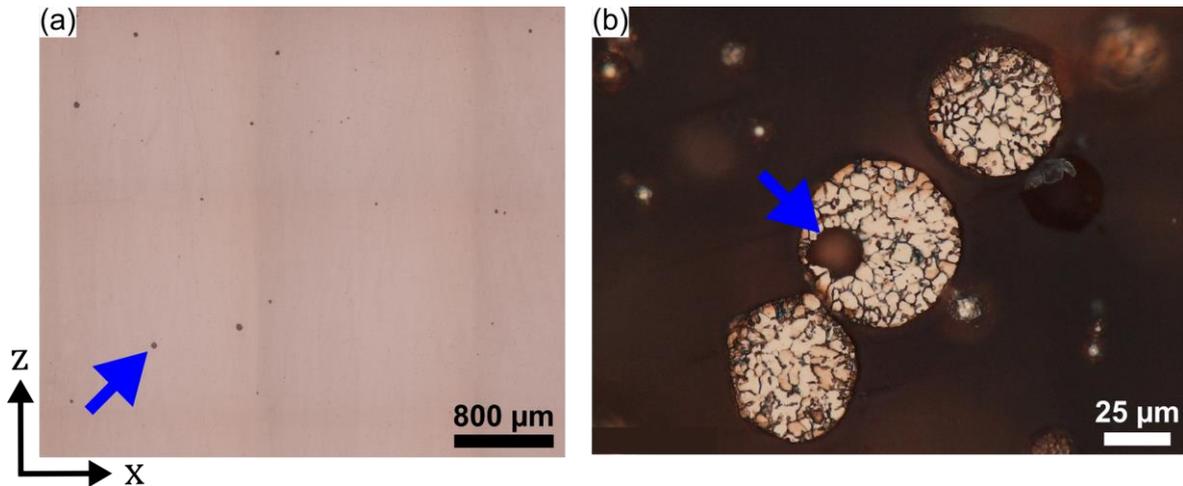

**Figure 4.** (a) Optical micrograph showing the presence of very few spherical pores pointed out by blue arrows in a 0°-sample. (x,y,z) is the coordinate system defined in **Figure 2**, x and y being the scanning directions, and z the build direction (BD). (b) 2D optical cross section illustrating the presence of pores in some powder particles.

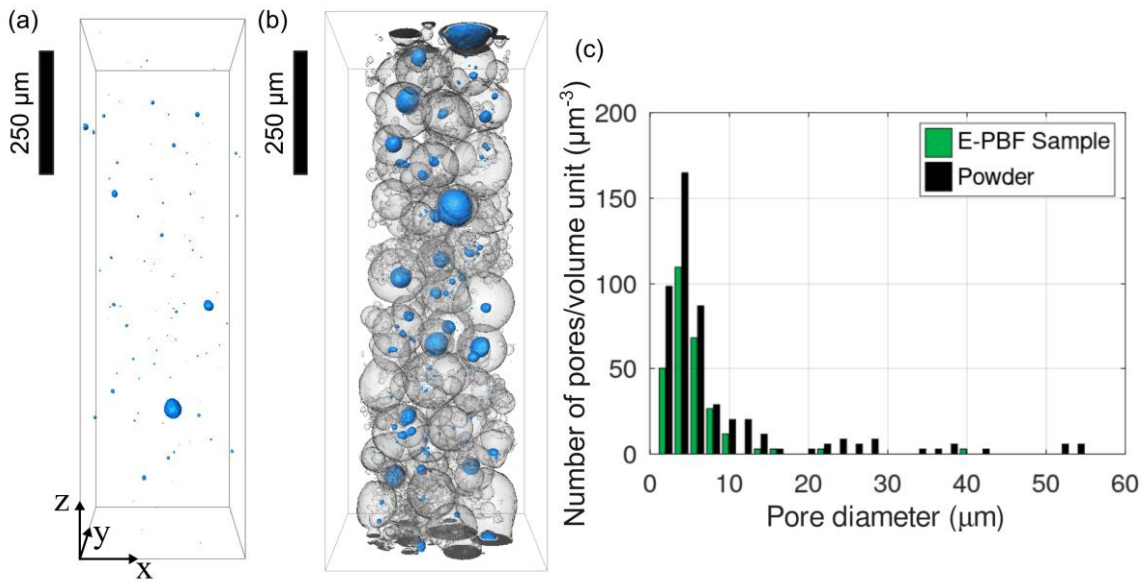

**Figure 5.** (a) 3D-image of pores observed in the as-built samples. (b) 3D-image of as-received powder particles (in light grey) with some containing pores (in blue). (x,y,z) is the coordinate system defined in **Figure 2**, x and y being the scanning directions, and z the build direction (BD). (c) Pore diameter distributions determined based on the 3D reconstructed images shown in (a) and (b).

Those spherical pores are thought to stem from the porosity contained in the initial powder particles. Here the powder was produced by gas atomization under Ar. During powder atomization, Ar can be trapped during solidification of the metal droplets leading to porous powder particles. This idea is supported by both optical observations conducted on powder particles cross sections (**Figure 4b**) and the 3D-reconstructed image of the as-received powder particles displayed in **Figure 5b**. In the 3D-image, 1086 powder particles are analyzed, the average diameter is calculated to be 72 μm, a value close to the $D_{v50}$ of the powder batch determined by laser granulometry. Interestingly, pores seem to be preferentially located within coarse particles. Over the 1086 powder particles analyzed, 261 contained pores. The internal porosity of the powder particles measured based on the 3D-image acquired by X-ray microtomography reaches 2.58 ± 0.03 %. This also allows to provide more detailed information regarding the pore morphology, size distribution and sphericity. Comparing the porosity in both as-fabricated samples and as-received powder particles suggests that a large fraction of the pores (more than 95 %) initially contained in the powder particles was suppressed during E-PBF. While melting under vacuum the powder particles with the high energy electron beam, most of these gas pores escape from the melt pool. Only a few of those spherical pores remained finally trapped in the melt pool and were left in the as-built samples as illustrated in **Figure 4a**. **Figure 5c** shows the diameter distribution of pores in the as-received powder particles as well as in the as-built E-PBF samples determined based on 3D-images. Both distributions are relatively similar for pores with a diameter < 10 μm. Pore diameters range from 5 to 50 μm, the majority being smaller than 10 μm. The latter observation is consistent with our assumption that those pores are inherited from the pores contained in the initial powder particles. The coarser pores present in the powder particles are no longer observed in the E-PBF samples suggesting that it is easier for large pores to escape from the melt pool during E-PBF because the upward buoyant force is higher for large pores as for small ones. Besides, one could also comment on the possible positive role of partial remelting of layer N-1 when processing layer N to decrease the density of such gas pores. Indeed, a gas pore might be retained in the material while melting layer N-1 (not enough time to escape from the molten bead) but this gas bubble has an additional chance to escape while melting layer N+1 if it is located closer to the surface or if it got bigger in case several pores had coarsened.

Better control of the gas atomization process or other atomization processing route should be considered if residual porosity (of the order of 0.05%) turns out to be an issue for the final performances of the built parts.

Another kind of defect can also be observed based on SE-SEM micrographs taken in the copper samples produced by E-PBF, see **Figure 6a**. These defects are rounded-shape pores aligned along the dendrites growth direction parallel to the build direction in this case. They are much smaller than the gas pores described previously, typically less than 1 μm in diameter and result from the coalescence of the dendrite secondary arms during the last stage of solidification [24] [25]. This second type of defects is qualified as micro-shrinkage pores and was not reported yet in copper parts built by E-PBF. **Figure 6b** schematically illustrates the mechanism leading to their formation. During the last stage of solidification, the growth of the dendrite secondary arms traps pockets of liquid metal. Pores are then created because of the volume contraction induced by the liquid to solid transformation. Note that, such defects are not observed in the 3D reconstruction because the resolution is not high enough to reliably capture them. Here, as the development of secondary arms appears limited, the solidification front could be qualified as cellular. Despite the fact that we are fabricating parts made of a high purity metal, one can still get a cellular/dendritic growth interface due to (i) a small constitutional undercooling (even about few degrees due to impurities solute rejection in the liquid) or (ii) a contribution from thermal undercooling.

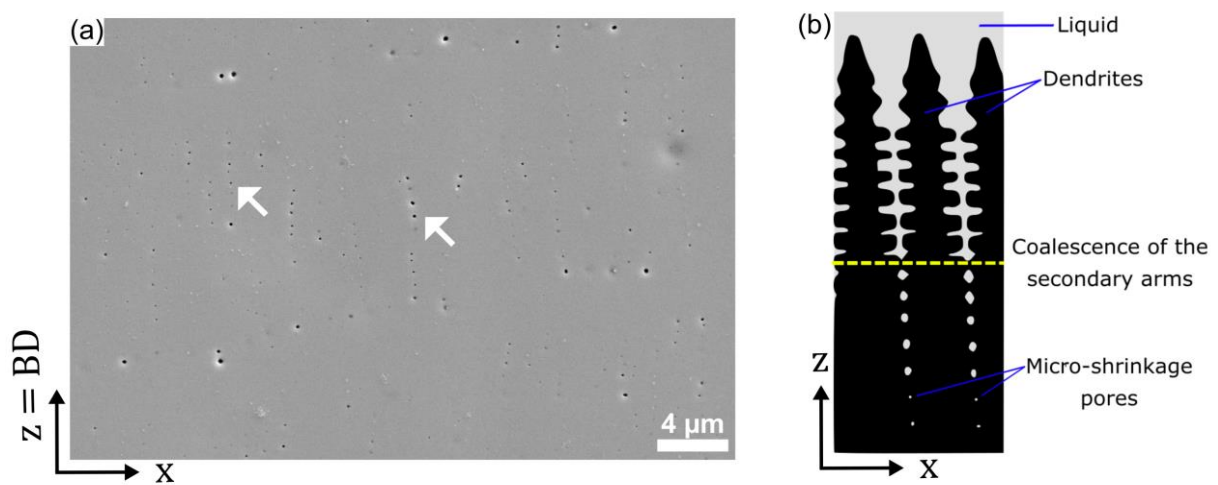

**Figure 6.** (a) SE-SEM micrograph revealing micro-shrinkage pores pointed out by white arrows. (b) Schematic illustration of the mechanism leading to micro-shrinkage due to the coalescence of the dendrite secondary arms during the last stages of solidification. (x,y,z) is the coordinate system defined in **Figure 2**, x and y being the scanning directions, and z the build direction (BD).

### 3.2. Microstructure

**Figure 7a-b** shows EBSD-IPF maps collected on a sample built vertically (0° orientation), in the xz-plane (plane containing the build direction) and xy-plane (plane perpendicular to the build direction) respectively. The microstructure is homogeneous along the build direction and consists of strongly textured <001> columnar grains elongated along the build direction, see **Figure 7a**. The average columnar grain width, measured using the intercept method is about 50 μm. Observations carried out by metallography in the topmost layer of the parts helps to rationalize microstructure formation during E-PBF. Grain and melt pool boundaries were revealed by chemical etching. For the selected processing parameters, the melt pool exhibits a relatively flat morphology (**Figure 7c**) with a depth about $70 \pm 8 \mu m$ which means that when one layer is melted, almost half of the previous layer is remelted. This promotes the epitaxial crystal growth leading to columnar grains going through several layers and elongated over several hundreds of microns. During melt pool solidification, the main direction of the thermal gradient is perpendicular to the solid-liquid interface. Thus grains grow perpendicularly to the melt pool boundaries. This is well-illustrated in **Figure 7c** by the orientation of the grain boundaries in the topmost layer. They are tilted in the direction of the thermal gradient at the edge of the melt pool while being well-aligned with the build direction at the center of the same melt pool. This peculiar feature is not observed in the underneath layers because those regions with tilted grain boundaries are remelted when the next layer or the adjacent track is melted.

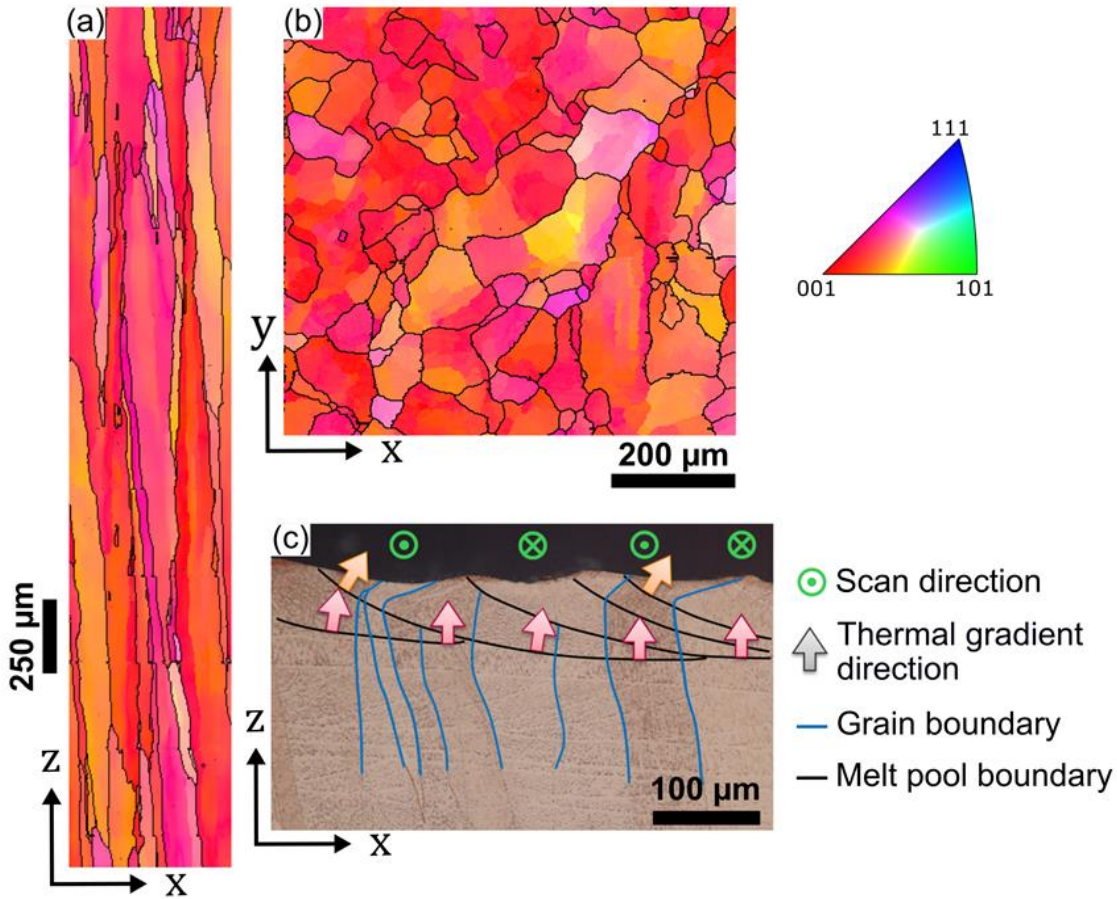

**Figure 7.** EBSD IPF maps of a 0°-sample giving an overview of the microstructure of the sample (a) in the xz-plane, and (b) in the xy-plane. (c) Optical micrograph of the topmost layer: melt pool boundaries highlighted with black lines and grain boundaries with blue lines. (x,y,z) is the coordinate system defined in **Figure 2**, x and y being the scanning directions, and z the build direction (BD).

<span style="color: red">**Figure 8a-c** show low-magnification optical micrographs taken in the xz-plane for the different samples built with different orientations. The microstructure consisting of columnar grains is found to be insensitive to the build orientation. The average columnar grain width is about 50 µm while the length of columnar grains can reach several millimeters. Note that the microstructure is found to be homogeneous along the build direction regardless the build orientation.</span>

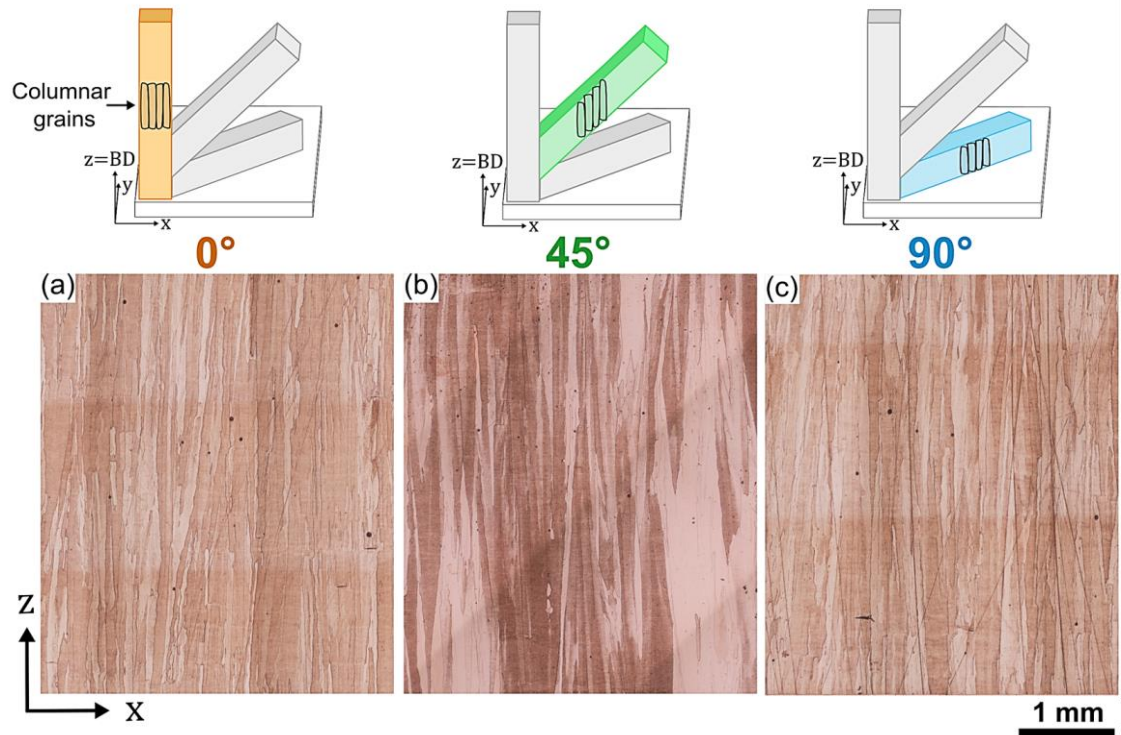

**Figure 8.** Optical micrographs taken in the xz-plane in samples built with the three build orientations: (a) 0°; (b) 45°; and (c) 90°.

**Figure 9a-c** show typical EBSD-IPF maps acquired in samples built with different orientation. Pole figures calculated based on large EBSD maps (not shown here) and associated to the three different build orientations are given in **Figure 9d-f**. Regardless the build orientation, samples have similar a very strong cube texture with a maximum intensity close to 20, see poles figures shown in **Figure 9d-f**. It means that the build direction (Z) is aligned with one of the <001> direction but the scanning directions x and y are also aligned with <001>-directions (cube texture). Such a strong cube texture with intensity of nearly 20 is a peculiar feature of pure Cu produced by E-PBF because pure Cu fabricated by traditional casting or forging does not show such an intense cube texture.

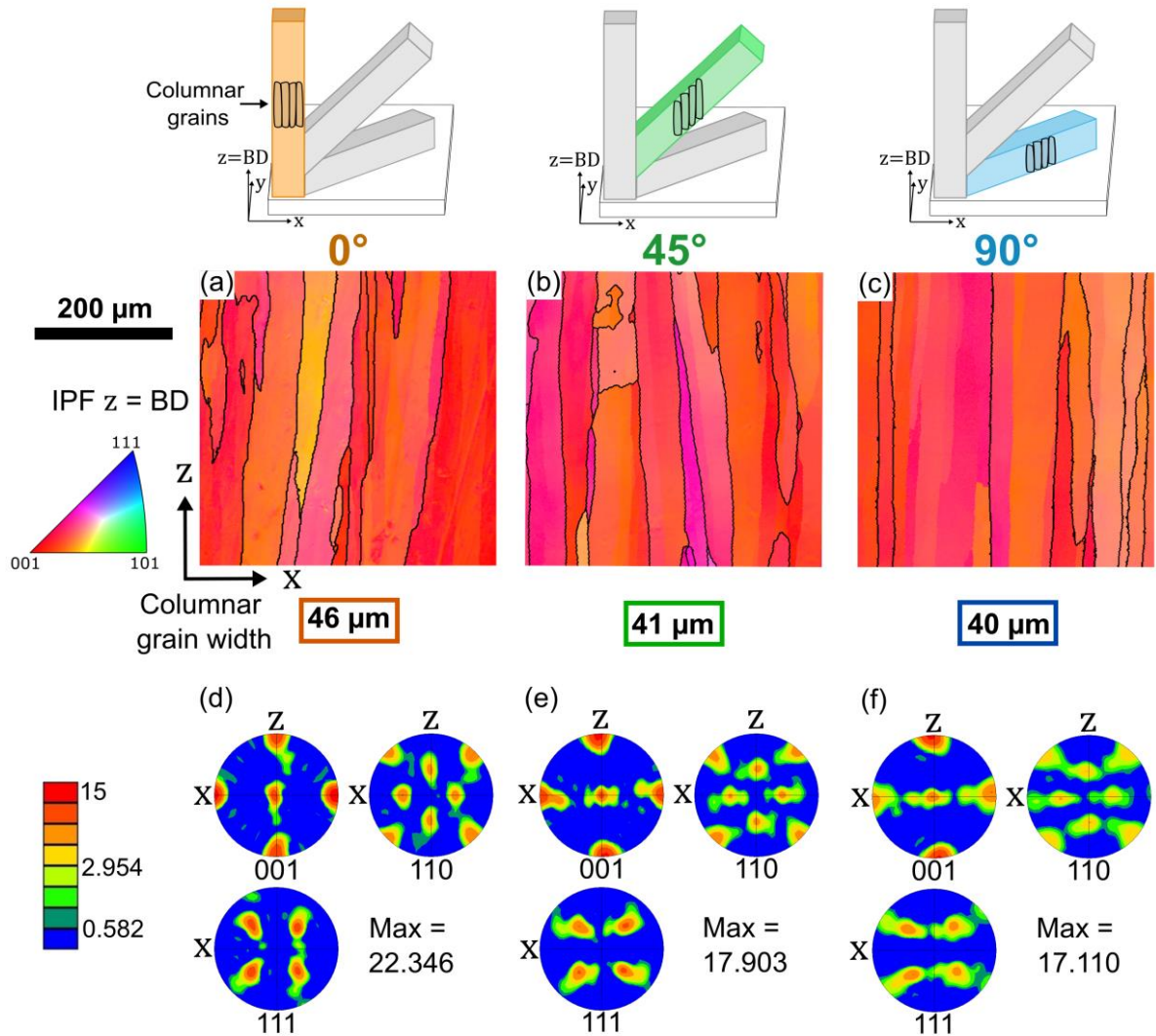

**Figure 9.** (a-c) Typical EBSD IPF-Z maps of the three build orientations, respectively 0, 45 and 90°. (d-f) (001, (110), and (111) Poles Figures determined on larger EBSD maps the 0,45, and 90°-oriented samples. (x,y,z) is the coordinate system defined in **Figure 2**, x and y being the scanning directions, and z the build direction (BD).

For comparison, the microstructure of wrought Cu-ETP annealed at 900°C for 15 min is presented in **Figure 10a** along with its texture represented with the help of pole figures (**Figure 10b**). The microstructure consists of equiaxed grains with a substantial fraction of twin boundaries. The average grain size is measured at 30 µm (twin boundaries included). This microstructure results from recrystallization and exhibits a weaker crystallographic texture than the E-PBF microstructure (maximum intensity about 5).

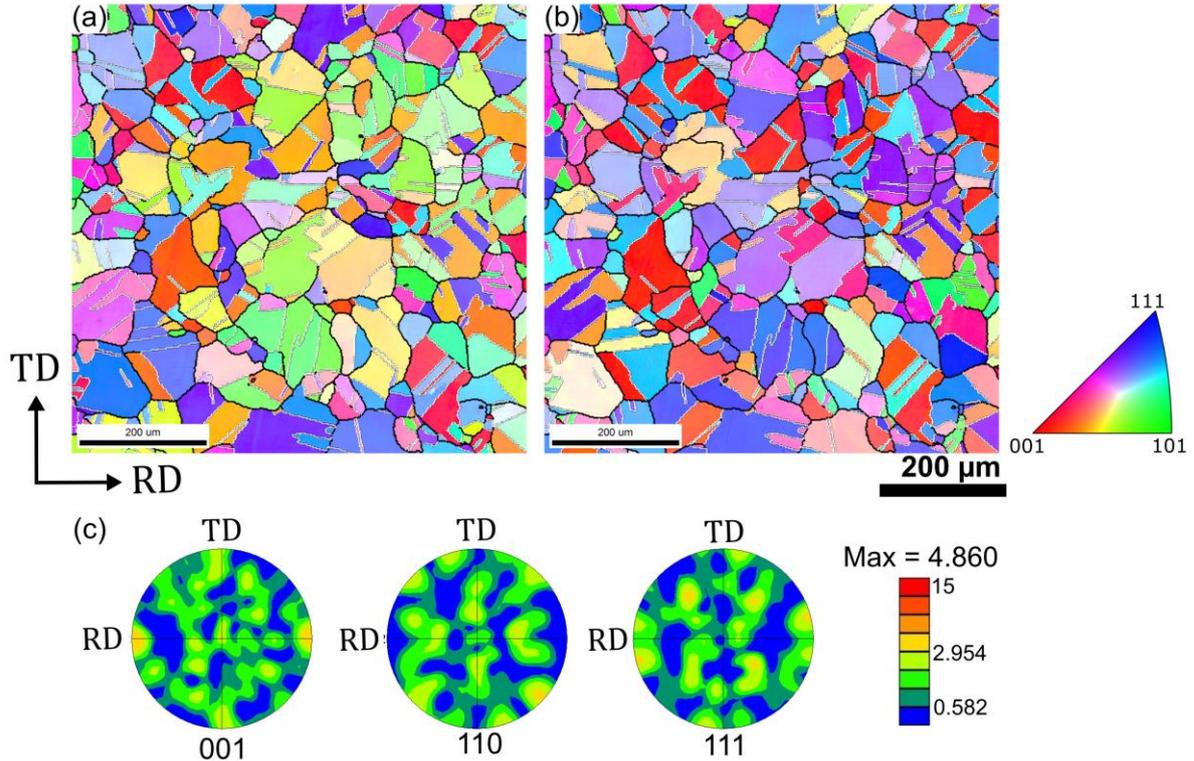

**Figure 10.** (a-b) EBSD IPF-TD and IPF-RD maps representing the grain orientation for the annealed Cu-ETP sample (cold-worked and annealed at 900°C for 15 min). RD, TD and ND are respectively the rolling direction, transverse direction, and normal direction of the metal sheet. Grain boundaries are represented with black lines and twin boundaries with gray lines. (c) (001), (110), (111) Pole figures corresponding to the area displayed in (a).

### 3.3. Electrical conductivity

Electrical conductivity measurements conducted on the three build orientations using both the Eddy current method (surface) and the 4-probes method (volume) are summarized in **Table 4**. The electrical conductivity of an annealed Cu-ETP sample was also measured using the Eddy current method for comparison.

| Sample build orientation (°) | 0° | 45° | 90° | Annealed Cu-ETP |
|---|---|---|---|---|
| **Surface $\sigma_{el}$ (%IACS)** <br> *Eddy current method* | 101.7 ± 0.6 | 102.0 ± 0.4 | 100.2 ± 3 | 100.5 ± 0.3 |
| **Bulk $\sigma_{el}$ (%IACS)** <br> *4-probes method* | 99.2 ± 0.4 | 100.0 ± 0.5 | 100.1 ± 1.1 | / |

**Table 4.** Electrical conductivity at 20°C of the bar samples for the three build orientations with two measurement methods: Eddy current (surface measure) and four probes method (bulk measurement). The measurement error is estimated by computing the standard deviation.

The electrical conductivity exceeds the standard requirement of the annealed Cu-ETP, i.e. 100 % IACS regardless the build orientation. This result is in agreement with the results reported by Gulschbauer *et al.* in [18] but better than the 97 % IACS reported in [2] or the 95% measured by Raab *et al.* [17], all reporting results on pure Cu fabricated by E-PBF. **Table 5** summarizes different results from the literature reporting the electrical conductivity of pure Cu processed by E-PBF or L-PBF for comparison. For now, the electrical conductivity of pure copper fabricated by L-PBF does not reach the 100 %IACS, the best value being 94% IACS and was reported recently in [10]. Our results also suggest that there is no influence of the build orientation on the electrical conductivity. One should keep in mind that the Eddy current method suffers from its low penetration depth for conductive materials (about 95 μm for pure copper). To measure the electrical conductivity in a significant volume representative of the bulk material, the four probes method was applied for the first time on specimens built by E-PBF. As reported in **Table 4**, the electrical conductivity of the three sample orientations reaches at least 99 % IACS. Two orientations (45 and 90°) meet the standard requirement of the annealed Cu-ETP, i.e. 100 % IACS while the 0°-specimens exceeds 99% IACS, most likely because of the presence of unexpected residual pores. Those measurements are in relatively good agreement with those performed using the Eddy current method but are thought to be more representative of the bulk.

| **E-PBF Processing conditions** | | | | | | | | **References** |
|---|---|---|---|---|---|---|---|---|
| Cu purity (%) | Preheating Temperature (°C) | Beam Power (W) | Beam speed (mm/s) | Angle with respect to the build direction (BD) | Relative Density (%) | Electrical conductivity (%IACS) | | |
| | | | | | | Eddy current | 4-probes | |
| 99.93 | 550 | 1200 | 3000 | 0°<br>45°<br>90° | 99.95 (Image analysis)<br>99.96 (XCT) | 101.7<br>102<br>100 | 99.2<br>100.0<br>100.1 | This work |
| 99.99 | 500-600 | - | - | 0° | - | 97 | - | [2] |
| 99.91 | 380-450 | - | - | 0° | 99.83 (Image analysis) | 96.2 | - | [17] |
| 99.95 | 530 | 850 | 1500 | 0°<br>45°<br>90° | > 99.5 (Image analysis)<br>-<br>- | 101.9<br>-<br>- | -<br>-<br>- | [18] |
| **L-PBF Processing conditions** | | | | | | | | |
| > 99.90 | - | 500 | 800 | 90° | 99.30 (Archimede) | - | 94 | [10] |
| > 99.90 | 80°C | 300 | 600 | 90° | 98.80 (Archimede) | 41 | - | [26] |
| **Cold Rolled + Annealed 900 °C/15 min** | | | | | | | | |
| Cu-ETP > 99.90 | - | - | - | Parallel to RD | 100% | 100.5 | - | This work |

**Table 5.** Summary of the electrical conductivity measurements found in the literature for pure copper processed by E-PBF and L-PBF. Results obtained in this work on a cold rolled and annealed Cu-ETP metal sheet are also reported for comparison.

### 3.4. Mechanical properties

To assess the influence of the build height on the mechanical properties, a hardness profile along the build direction was performed on the 0°-sample. A constant hardness value of HV0.5 = 50 ± 3 is found from the bottom to the top of the sample. This result indicates that there is no influence of the build height on the hardness. We recall that during E-PBF each layer is preheated before being selectively melted. This means that the successive layers built undergo an *in situ* annealing which duration depends on the height of the samples. Here as there is no variation of hardness along the sample height, it suggests that the shortest annealing time (last layers built) is enough to achieve the same hardness as the one measured in the first few layers that underwent the longest annealing time (first layers built). The hardness measured for each build orientation is given in **Table 6**. Regardless of the build orientation a hardness about 50 HV0.5 is measured. This is close though slightly higher than the hardness measured on the annealed cold-worked Cu-ETP.

| Sample build orientation (°) | 0° | 45° | 90° | Cu-ETP | |
|---|---|---|---|---|---|
| | | | | Cold-worked | Cold-worked + Annealed (900°C/15min) |
| **Hardness HV0.5 (xz-plane)** | 51 ± 1 | 50 ± 3 | 51 ± 2 | 97 ± 3 | 46 ± 1 |

**Table 6.** Vickers hardness values of bar samples for the three build orientations, the measurements are performed in the xz-plane with z the build direction. Their values are compared with the ones measured on Cu-ETP samples.

**Figure 11a** summarizes the true tensile mechanical response of the three build orientations. Good reproducibility of the tests should be emphasized. The mechanical properties extracted from the tensile responses shown in **Figure 11a** are given in **Table 7**, except the elongation at fracture that was extracted from the engineering stress-strain curves. Small differences in yield strength between the three build orientations are observed. Interestingly they show similar values as the annealed Cu-ETP. However, the true ultimate tensile strength (UTS) is affected by the build orientation. The 0°-samples exhibit a significantly lower UTS than the 45 and 90°-samples, and the annealed Cu-ETP samples. The UTS of the 0°-orientation is roughly about 70% of the minimum value required for annealed Cu-ETP (300 MPa as indicated in the EN 13601 standard). **Table 8** summarizes the results of different studies reporting mechanical properties of pure Cu fabricated by E-PBF and L-PBF. Our results (hardness, yield strength, elongation to failure, and UTS) on the 0°-specimens built by E-PBF are in relatively good agreement with those of Gulschbauer *et al.* [18] though differences exist for the two other build orientations (45 and 90°). Gulschbauer *et al.* [19] reported more recently higher values for the yield strength for all three build orientations, this is likely due to a lower build temperature (440 vs. 550°C in this work). In the literature, we did not find a study dedicated to pure copper processed by L-PBF and that aimed at comparing the mechanical properties of different build orientations. However, the few data available in the literature show that the hardness and yield strength of pure copper processed by L-PBF are higher than the ones measured on E-PBF samples that goes along with a reduction of the ductility, see **Table 8**. This is not surprising given that during E-PBF, contrary to during L-PBF, parts undergo an in situ annealing caused by high build temperature.

| Samples | 0° | 45° | 90° | Annealed Cu-ETP |
|---|---|---|---|---|
| **True UTS (MPa)** | 216 ± 2 | 304 ± 9 | 263 ± 14 | 309 ± 3 |
| **Yield Strength (MPa)** | 66 ± 2 | 67 ± 1 | 60 ± 1 | 67 ± 2 |
| **Elongation to failure (%)** | 52 ± 4 | 52 ± 3 | 47 ± 3 | 54 ± 3 |
| **Reduction of area (%RA)** | 90 | 80 | 60 | 75 |

**Table 7.** Mechanical properties estimated from the true tensile curves (except for the elongation at fracture measured on the engineering curves and the reduction of area determined based on the fracture surface). Tests have been repeated 3 times and only the average value is given here.

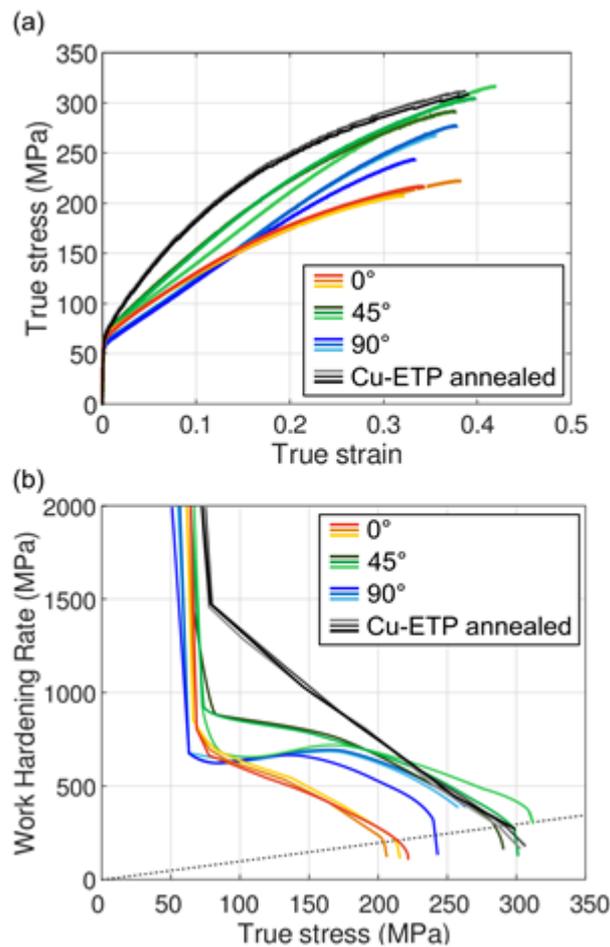

**Figure 11.** Tensile results for the three kinds of samples produced by E-PBF and compared with annealed Cu-ETP: (a) True stress / True strain tensile response until the onset of necking (uniform deformation only); (b) Work-hardening rate as a function of true stress with the Considere criterion in dotted ray lines.

| **E-PBF Processing conditions** | | | | | | | | | **References** |
|---|---|---|---|---|---|---|---|---|---|
| Cu purity (%) | Preheating Temperature (°C) | Beam Power (W) | Beam speed (mm/s) | Angle with respect to the build direction (BD) | Vickers Hardness (HV) | Yield strength (MPa) | UTS (MPa) | Elongation to failure (%) | |
| 99.93 | 550 | 1200 | 3000 | 0° | 51 | 66 | 144 | 52 | |
| | | | | 45° | 50 | 67 | 206 | 52 | This work |
| | | | | 90° | 51 | 60 | 184 | 47 | |
| 99.80 | 550 | 600 | 80 | - | 83 | - | - | - | [13] |
| 99.99 | 500-600 | - | - | 0° | - | 76 | - | - | [2] |
| 99.95 | 530 | 450 | 500 | 0° | 46-48 | 78 | 177 | 59.3 | |
| | | | | 45° | - | 87.5 | 158 | 14.3 | [18] |
| | | | | 90° | - | Cracked | Cracked | Cracked | |
| - | 440 | 450 | 500 | 0° | - | 125 | 181 | 60 | |
| | | | | 45° | - | 125 | 205 | 51 | [19] |
| | | | | 90° | - | 150 | 232 | 50 | |
| **L-PBF Processing conditions** | | | | | | | | | |
| > 99.90 | - | 500 | 800 | 90° | 66 | 122 | 211 | 43 | [10] |
| > 99.90 | 80 | 200 | 400 | 90° | 84 | 187 | 248 | 9 | [26] |
| **Cold Rolled + Annealed 900 °C/15 min** | | | | | | | | | |
| Cu-ETP > 99.90 | - | - | - | Parallel to RD | 46 | 67 | 214 | 54 | This work |

**Table 8.** Summary of the mechanical properties found in the literature for pure Cu processed by E-PBF and L-PBF. Mechanical properties measured in this work on a cold worked and annealed Cu-ETP metal sheet are also reported for comparison. Yield strength and UTS are the engineering values.

The evolution of the work-hardening rate along with the Considere criterion (in dotted gray line) are plotted in **Figure 11b** for the three samples built with different orientations as well as for the cold-worked annealed specimens. The 0°-specimens exhibit a two-stages work-hardening. In stage I, the work hardening rate suddenly drops ($\sigma < 75$ MPa). Then, during stage II, it continuously decreases. For the 45°-specimens (except for one), three stages are observed. Similar to the 0°-specimens, stage I consists of a sharp drop in the work-hardening rate. Then, a plateau appears where the work-hardening rate smoothly declines (stage II). In stage III, the work-hardening rate decreases with approximately the same behaviour as the 0°-specimens. For the 90°-specimens, the work-hardening curve also exhibits three stages. The first (I) and last (III) stages are relatively similar to the ones observed in the 45°-specimens. However, in stage II, a rise of the work-hardening rate with the increase of the true stress is observed. For the annealed Cu-ETP samples, only two stages can be distinguished in the work-hardening rate curve as observed for the 0°-specimens. It first decreases sharply, then it declines uniformly at a lower rate. The regime change occurs at a higher rate than for the bar samples produced by E-PBF. Similar two-stages work-hardening behaviour is reported in the literature for annealed pure copper, see Sinclair *et al.* for instance [27] [28].

The reduction area (*%RA*) defined as follows $\%RA = 100.(A_i - A_f/A_i)$, $A_i$ and $A_f$ being respectively the initial and final cross-sectional area of the tensile specimens were also estimated. The final cross-sectional area ($A_f$) was measured based on the SEM images of the fracture surface for each build orientation along with the one of a cold-worked annealed Cu-ETP. The average results (calculated for 3 specimens in each case) are reported in **Table 7**. The 0 and 45°-specimens have the highest *%RA* values with respectively 90 and 80%. These values are similar to the reported values for Cu-OF by Jenkins *et al.* [29] : *%RA* = 85-88%. The 90°-specimens stand out with a lower *%RA*, close to 60%. Finally, the *%RA* of the cold-worked annealed Cu-ETP is measured to be about 75%.

## 4. Discussion

### 4.1. Electrical conductivity

In metals, the electrical conductivity is defined as the ability of a flow of electric charges, i.e. electrons to go through the material. The last electronic orbital of copper exhibits a single electron; this electron is responsible for the high conductivity of copper. Here, all specimens built by E-PBF with various orientations exhibit an excellent electrical conductivity, higher than 100 % IACS (58 MS/m), see **Table 4**. This result is discussed here in the light of the different factors expected to affect the electrical conductivity. Among those factors, porosity, impurity atoms especially when those are in solid solution [30] [31], vacancies [32], grain boundaries [33] [34], and dislocations [35] [31] are known to be detrimental to the electrical conductivity [30] [31].

Porosity is known to lead to a knock-down of the electrical conductivity when its content becomes higher than 1%, see [36] [37] [38]. In the present work, the processing conditions were optimized in a preliminary study and allow high relative densities to be reliably achieved regardless of the build orientation. The relative density was estimated by image analysis and values exceeding 99.9% were measured. The latter result was confirmed by a 3D analysis based on X-ray microtomography where the relative density was found to be 99.96%. The residual porosity (0.04%) is attributed to few spherical gas pores resulting from pores initially trapped within powder particles and that did not successfully escape from the melt pool during E-PBF. Given the very low porosity content measured in this work, it is not expected to have an effect on the electrical conductivity.

In pure copper, a few ppm of impurity atoms in solid solution can drastically decrease the electrical conductivity. The higher the element concentration, the lower the electrical conductivity. Elements such as Fe, P or Si are the most harmful for the electrical properties [30]. For instance, 0.004 wt% of P is expected to decrease the conductivity down to 97% IACS [30]. For similar concentrations, in Fe and Si, a loss of electrical conductivity of roughly 3% IACS is also expected. Thus, the concentration in Fe, Si and P along with the O and H contents were measured in the as-received powder as well as in the parts built by E-PBF, see **Table 9**. The most detrimental elements were systematically found below 0.005 %wt but their exact concentration could not be measured more accurately. However, given that electrical conductivities equal or higher than 100%IACS were achieved, it suggests that their concentrations are below the critical concentration that would lead to a few percent loss of the electrical conductivity. An increase of 40 ppm in oxygen: from 110 ppm in the powder to 150 ppm in the parts was detected while a substantial enrichment of H was measured: from 1 ppm in the powder to 20 ppm in the parts. This suggests that those two interstitial elements would need to be closely monitored, especially if one plans to study powder recycling.

|  | **Fe** | **Si** | **P** | **O** | **H** |
|---|---|---|---|---|---|
| **Cu as-received powder** | < 0.005 | < 0.005 | < 0.005 | 0.011 | 0.0001 |
| **Cu fabricated by E-PBF** | < 0.005 | < 0.005 | < 0.002 | 0.015 | 0.002 |

**Table 9.** Concentrations in Fe, Si, P, O and H given in wt% in the as-received powder bath and in a sample built by E-PBF.

A high density of vacancies can also have an effect on the electrical conductivity because the presence of vacancies induces a local distortion of the lattice resulting in scattering the conduction of electrons that further leads to a decrease of the measured electrical conductivity at the macroscopic scale. Overhauser *et al.* calculated the resistivity associated with one atomic percent of vacancies in copper and found 1.5 $\mu\Omega$.cm [32]. The density of vacancies varies with the thermo-mechanical history of the material. For instance, very high cooling rates might prevent diffusion mechanisms and extra vacancies can be trapped. On the opposite, annealing allows diffusion and therefore can result in a reduction of the density of vacancies. Unlike L-PBF, the powder bed is maintained at high temperature throughout the build, about 550°C in the present work. It means that the parts undergo an *in situ* heat treatment during which solid state diffusion can happen. As a result, a low density of vacancies is suspected, suggesting no effect on the electrical conductivity.

The *in situ* annealing induced by the preheating stage is also thought to play a key role by relieving residual stresses and more importantly by acting as a recovery treatment that contributes to decrease the dislocation density in the as-fabricated parts. This idea was supported by our microhardness measurements, see **Table 6**. A Vickers microhardness of about 50 HV0.5 was measured for the three build orientations, such value is very close to the value obtained in a cold-worked Cu-ETP sample annealed at 900°C for 15min indicating that the dislocation density is low in the as-built parts and therefore should not impact the electrical conductivity [31].

Finally, the grain boundary density which can be reflected by the grain size is also a factor that could affect the electrical conductivity. However, the effect of grain size on the conductivity starts to become significant only for grain size typically < 1 $\mu$m [33] [39]. Here, with columnar grain width of the order of 50$\mu$m, the grain boundary density is not high enough to affect the electrical conductivity. Also one should note that in the case of the bulk measurements (4-probes method) the grain boundaries are oriented differently with respect to the path followed by electrons and this does not result in differences in electrical conductivity.

To summarize, the high electrical conductivity measured on the copper samples produced by E-PBF being as good as that of a Cu-ETP cold-worked and annealed sample (see **Table 4**), it suggests that all those factors, namely pores, impurity atoms in solid solution, vacancies, grain boundaries, and dislocation density are all kept below their critical value that would affect the electrical conductivity.

### 4.2. Mechanical properties

Interestingly, the yield strengths measured for the three build orientations are very close: 66, 67, and 60 MPa for respectively the 0, 45, and 90°-orientation. Here, these very similar yield strengths are discussed in the light of the different strengthening mechanisms. The raw material being a high purity metal, solid-solution or precipitation strengthening mechanisms do not contribute to the strength of the material. Only grain boundary strengthening (Hall-Petch effect) or forest hardening can thus contribute to the strength of pure copper.

Miura *et al.* [40] and Hansen *et al.* [41] determined Hall-Petch laws ($\sigma_y = \sigma_0 + k.d^{-0,5}$) for polycrystalline annealed pure copper. Note that the Hall-Petch parameters were identified for grain sizes ranging from $d = 8$ to 40 μm in [40] ($\sigma_0 = 7.9$ MPa and $k = 0.13$ MPa.m$^{0.5}$) and from $d = 10$ to 150 μm in [41] ($\sigma_0 = 20$ MPa and $k = 0.16$ MPa.m$^{0.5}$). The morphology of the grain does not vary when comparing the three build orientations (columnar grains). More importantly the characteristic length of the grains taken here as the average columnar grain width measured using the intercept method was found to be of the same order: 46, 41 and $40 \pm 5$ μm for the 0, 45 and 90°-specimens respectively. Using the Hall-Petch relationships found in the literature [40] [41], such slight differences in grain size should not result in a significant difference of yield strength. This is consistent with the literature showing that the effect of the grain size on the yield strength starts to become significant for grain sizes < 10 μm, see [39]. One could question the use of the Hall-Petch law because it assumes grains with an equiaxed morphology whereas here we have highly elongated grains (some grains being > 1 mm in length) loaded along different directions depending on the build orientation. What matters when considering grain boundary strengthening is the distance over which dislocations can glide before facing a grain boundary. As dislocations move in specific planes along given directions ({111}-<110> slip systems for a FCC-material such as pure Cu), we think that grain boundary strengthening will be primarily governed by the minimum characteristic length, i.e. the columnar grain width here. The latter consideration likely helps to rationalize why all three build orientations have nearly the same yield strength.

In addition, the absence of intragranular misorientation in the EBSD maps combined with a hardness of about 50 HV0.5 measured for the three build orientations and very close to the hardness of annealed pure copper (46 HV0.5, **Table 6**) suggest that forest hardening is not contributing neither.

While the yield strengths do not change when comparing the three build orientations, the work-hardening of the samples shows some differences : the work-hardening of the 0°-specimens differing from the ones of the 45 and 90°-specimens, see **Figure 11a-b**. To our knowledge, such difference of work-hardening between different build orientations has not been investigated yet and was not highlighted in previous works studying the mechanical properties of pure copper fabricated by E-PBF, see e.g. [18], [19]. The grain size was found to have an effect on the work-hardening of copper, see [27]. However, this grain size dependence of the work-hardening was limited to samples with grain size of the order of 10 μm, this effect vanishing for coarser grain microstructures [27]. As the grain size is relatively coarse (columnar grains roughly 40μm wide) and can be considered constant between the samples fabricated by E-PBF with three build orientations, we do not believe that the differences of work-hardening result from a difference of grain size. This difference of work-hardening was rather attributed to a difference of crystallographic texture. A strong cube texture with an intensity close to 20 was found in the samples built by E-PBF regardless the build orientation, see **Figure 9**. However, the cube texture was not loaded along the same axis: in the 0°-specimens the tensile direction was aligned with a <001>-direction whereas the tensile direction was rather aligned with a <110>-direction for both the 45 and 90°-specimens, as illustrated in **Figure 12a-c**.

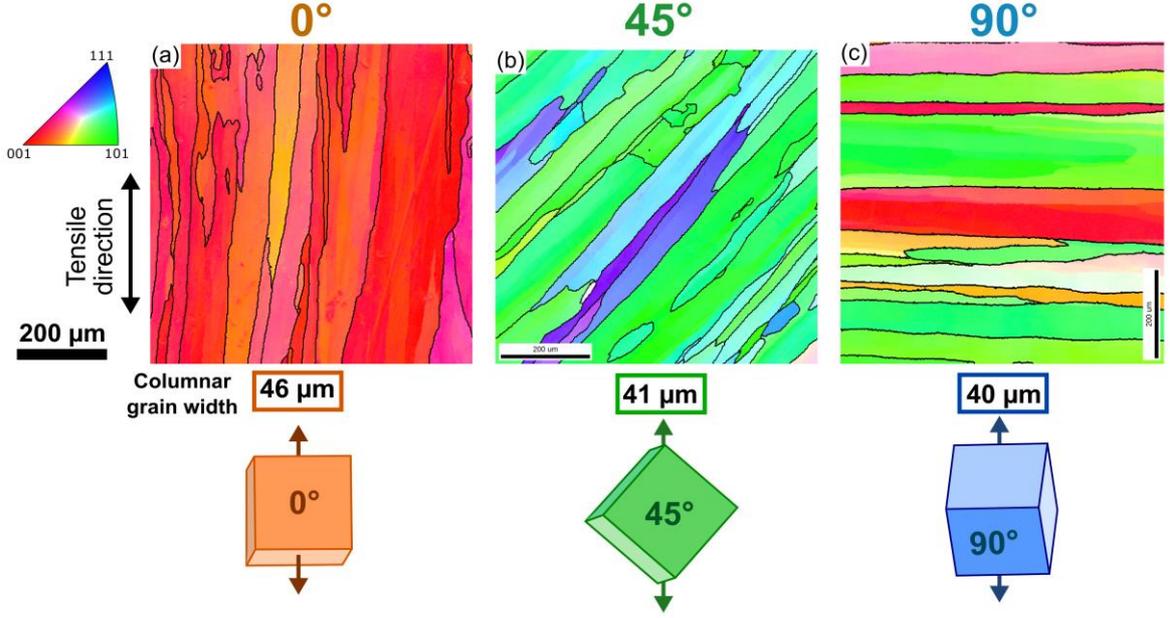

**Figure 12.** EBSD-IPF maps representing the grain orientation along the tensile direction for (a) 0°-specimens; (b) 45°-specimens; and (c) 90°-specimens.

To further clarify the effect of the crystallographic texture on the work-hardening of the samples produced by E-PBF with various build orientations, we used a mean-field crystal plasticity model (VPSC: Visco-Plastic Self Consistent) model, initially developed to predict the strain-stress response of textured materials, see examples in [42] [43]. In the VPSC model, each grain is considered as a visco-plastic anisotropic ellipsoidal inclusion embedded in a visco-plastic Homogeneous Effective Medium (HEM) [42]. The VPSC model utilizes the formalism suggested by Eshelby: the stress and strain rates are uniform inside the inclusion but differ from the macroscopic (average) values. In this study, the Los Alamos VPSC model (version 7d) has been employed to simulate the tensile stress-strain response of the bar samples built with three different orientations, namely 0°, 45°, and 90°. Calculations were performed using a texture file containing about 200 grain orientations weighted by their volume fraction based on large EBSD maps acquired to determine the representative texture of the 0, 45 and 90°-specimens. The strain rate was imposed along the macroscopic tensile direction while its transverse components were set as unknowns and incremental straining steps of 0.01 were used. In our calculation, a strain rate sensitivity exponent $n = 20$ was used to describe the rate dependence of slip systems, and the interaction parameter from the Eshelby's inclusion formalism is assigned to be $n_{eff} = 10$ (intermediate between Taylor upper bound: $n_{eff} \rightarrow 0$, and Sachs: $n_{eff} \rightarrow \infty$ lower bound) [42]. This is a typical value used by many authors to calculate the mechanical response of FCC-materials, see e.g. [42]. A single deformation mode was considered, namely slip in the {111} planes along the <110> directions. A Voce-type hardening law was chosen to describe the hardening of individual slip system, see equation (4).

$$\tau^S = \tau_0 + (\tau_1 + \theta_1\Gamma)\left(1 - \exp\left[-\theta_0\Gamma / \tau_1\right]\right) \tag{4}$$

where $\tau^S$ is the threshold stress which describes the resistance for the activation of particular deformation modes which usually evolves with deformation due to strain hardening and $\Gamma$ is the accumulated shear strain in the grain. $\tau_0$ and $\tau_1$ are the initial and back-extrapolated increase in the threshold stress, while $\theta_0$ and $\theta_1$ are the initial and final slope of the hardening curve. Note that we deliberately kept the model as simple as possible by making 'self' and 'latent' hardening indistinguishable. In other words, it means that we do not account for the obstacles that new dislocations associated to a given slip system represent for the propagation of dislocations in another slip system.

The four parameters $\tau_0$, $\tau_1$, $\theta_0$, and $\theta_1$ involved in the equation governing the hardening law of individual slip system were first optimized to fit the tensile response of the 0°-samples (the orientation showing a two stage work-hardening). The optimized set of parameters are summarized in **Table 10**.

| $\tau_0$ (MPa) | $\tau_1$ (MPa) | $\theta_0$ | $\theta_1$ |
|:---:|:---:|:---:|:---:|
| 28 | 140 | 85 | 5 |

**Table 10.** Optimized set of parameters describing the hardening of individual slip system used in the VPSC calculations.

This set of optimized parameters was then applied to predict the tensile response of both the 45 and 90°-specimens. The exact same VPSC calculations were run but the input texture file was modified using the EBSD measurements and the loading direction was chosen so as to reproduce the experimental loading conditions. The VPSC predictions of the tensile response are plotted along with the experimental data for each build orientation, see **Figure 13a-c**.

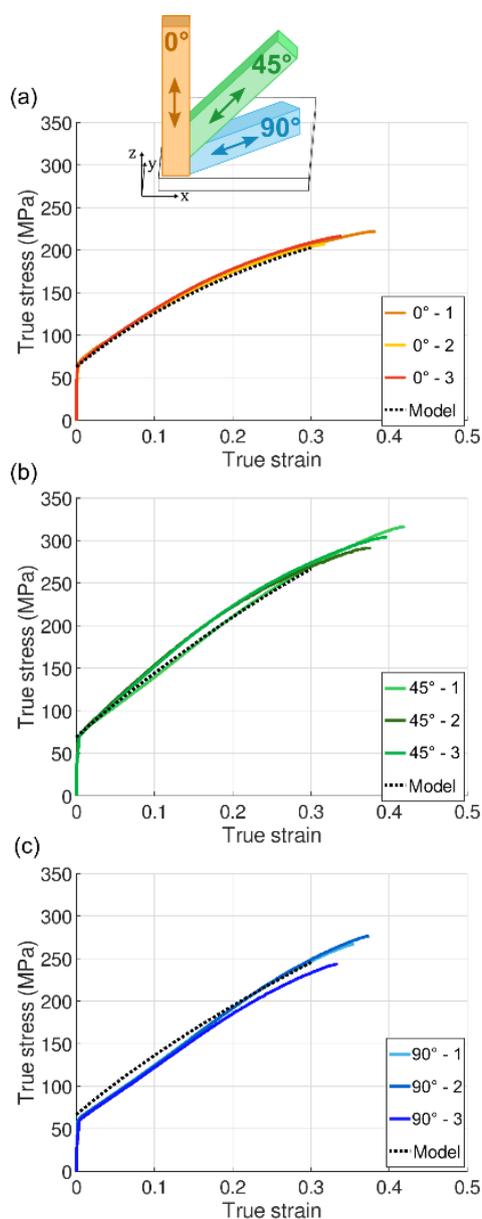

**Figure 13.** Comparison of the experimental tensile response with the ones predicted using the VPSC model using the parameters summarized in **Table 6** for (a) the 0°-specimens; (b) the 45°-specimens; and (c) the 90°-specimens.

Overall, a good agreement is found between the tensile response of the three different build orientations and the corresponding VPSC predictions. The increase of flow stress defined as $\Delta\sigma(\varepsilon) = \sigma(\varepsilon) - \sigma_y$, $\sigma(\varepsilon)$ being the flow stress for a given tensile strain $\varepsilon$, and $\sigma_y$ the yield strength, can be used to evaluate the work-hardening. Thus, $\Delta\sigma(\varepsilon)$ have been estimated for $\varepsilon = 0.1$; 0.2 and 0.3 based on the experimental tensile cruves as well as on the predicted tensile responses. The comparison is given in **Table 11** and reveals a relatively good agreement between experiments and predictions.

| | | *Δσ(ε = 0.1)* MPa | *Δσ(ε = 0.2)* MPa | *Δσ(ε = 0.3)* MPa |
|---|---|---|---|---|
| **Experiments** | 0° | 63 | 109 | 138 |
| | 45° | 73 | 144 | 203 |
| | 90° | 65 | 132 | 187 |
| **VPSC predictions** | 0° | 63 | 109 | 140 |
| | 45° | 75 | 141 | 198 |
| | 90° | 69 | 128 | 179 |

**Table 11.** Estimations of the increase of flow stress at different macroscopic tensile strains ofr the three different build orientations.

The results are consistent with our initial assumption that the difference in work-hardening between the 0°-specimens on the one hand, and the 45 and 90°-orientations on the other hand, can be accounted for a difference of crystallographic texture, the cube texture being loaded along different directions. The cube texture of the 45 and 90°-specimens were loaded along a similar crystallographic direction, this is why they roughly show a similar work-hardening. On the opposite, the 0°-specimens were loaded along a different direction and therefore has a different work-hardening. Though our predictions capture the main trends, one should note that the different work-hardening regimes identified in **Figure 11b**, typically the different work-hardening stages evidenced for both 45° and 90°-orientations. We believe that this might be because we used a simple VPSC formulation that does not take into account the interaction strength and the anisotropic softening between slip systems as suggested in [28]. Interestingly, Bacroix *et al.* [28] showed that using a Taylor type model taking into account a dislocation-based hardening law including anisotropic hardening and softening terms allows the different work-hardening regimes to be much better described.

## 5. Conclusion

High purity copper (99.93 % wt) was successfully fabricated by E-PBF achieving a density >99.95% with very few residual pores: gas pores resulting from powder atomization and microshrinkage during the last stages of solidification. Three different build orientations were produced to investigate the effect of build orientation on both the electrical and the mechanical properties. The microstructure consists of coarse columnar grains elongated along the build direction with a strong cube texture resulting from epitaxial grain growth. The same microstructure and microtexture were found regardless the build orientation. The relationships between the microstructure and the properties have been clarified and compared with a cold-worked and annealed Cu-ETP. The electrical conductivity measured successively by the Eddy current method and the four probes method is found to be higher than 100% IACS regardless of the build orientation. This high conductivity was rationalized by a very limited presence of pores, the absence of detrimental solutes as well as the presence of coarse columnar grains and a low dislocation density caused by the preheating step which acts as a recovery heat treatment. The mechanical properties, namely the hardness ($\approx 50$ HV$_{0.5}$), yield strength ($\approx 65$ MPa), and the elongation to failure ($\approx 50\%$), are very close for the three build orientations and similar to the properties measured on cold-worked and annealed Cu-ETP specimens. The only differences found between the different build orientations are the work-hardening and the ultimate tensile strength. The difference of work-hardening was accounted for a texture effect because even if the three build orientations show a similar cube texture, the latter was loaded in different directions resulting in different work-hardening. This conclusion was further supported by crystal plasticity calculations using the VPSC model.

**Ackowledgements**

Schneider-Electric is gratefully acknowledged for the financial support of the present study. The authors are grateful to the LERMPS (Laboratoire d'Etudes sur les Matériaux, les Procédés, et les Surfaces) for the production of the powder used in this work. The present study was carried out in the framework of the FUI-Ambition research program. This work has benefited from the characterization equipment of the Grenoble INP - CMTC platform and the E-PBF machine financially supported by the Centre of Excellence of Multifunctional Architectured Materials "CEMAM" n°ANR-10-LABX-44-01 funded by the Investments for the Future programme.

10.1016/j.msea.2020.139615.